\documentclass[10pt]{iopart}
\pdfoutput=1

\expandafter\let\csname equation*\endcsname\relax
\expandafter\let\csname endequation*\endcsname\relax
\usepackage{hyperref}

\usepackage[english]{babel}
\usepackage{amsmath}
\usepackage{amsfonts}
\usepackage{amssymb}
\usepackage{color}
\usepackage{epsfig}
\usepackage{graphicx}
\usepackage{bm}
\usepackage{mathtools}
\usepackage{bbold}
\usepackage{times}
\usepackage{cancel}

\allowdisplaybreaks 
\newcommand{\E}{\mbox{$\mathsf E$}}

\newcommand{\abs}[1]{\left| #1 \right|} %
\newcommand{\bra}[1]{\langle #1 |} \newcommand{\ket}[1]{| #1 \rangle}
\newcommand{\scp}[2]{\langle #1 | #2 \rangle}
\newcommand{\braket}[3]{\langle #1 | #2 | #3 \rangle}

\newcommand{\BigO}[1]{\ensuremath{\operatorname{O}\left(#1\right)}}
\newcommand{\Span}{\mathrm{span}} %

\newcommand{\trcond}{\mathop{\mathrm{tr}_\mathrm{cond}}\limits}
\newcommand{\trnorm}{\mathop{\mathrm{tr}_\mathrm{norm}}\limits}

\definecolor{myred}{RGB}{168,5,14}
\definecolor{myblue}{RGB}{13,13,255}
\definecolor{mylightblue}{RGB}{120,120,255}
\definecolor{mygreen}{RGB}{20,150,20}
\definecolor{mygray}{gray}{0.5}

\definecolor{cpcolor}{RGB}{12,30,255}

\begin{document}

\title[Finite temperature quantum condensations in the space of
states: General Proof]{Finite temperature quantum condensations in the
  space of states: General Proof}

\author{Massimo Ostilli} \address{Instituto de F\'isica,
  Universidade Federal da Bahia, Salvador, 40210-340, BA, Brazil}
\author{Carlo Presilla} \address{Dipartimento di Fisica, Sapienza
  Universit\`a di Roma, Piazzale A. Moro 2, Roma 00185, Italy}
\address{Istituto Nazionale di Fisica Nucleare, Sezione di Roma 1,
  Roma 00185, Italy}

\vspace{10pt}
\begin{indented}
\item[]\today
\end{indented}

\begin{abstract}
  We formalize and prove the extension to finite temperature of a
  class of quantum phase transitions, acting as condensations in the
  space of states, recently introduced and discussed at zero
  temperature~(Ostilli and Presilla 2021 \textit{J. Phys. A:
    Math. Theor.} \textbf{54} 055005).  In details, we find that if,
  for a quantum system at canonical thermal equilibrium, one can find
  a partition of its Hilbert space $\mathcal{H}$ into two subspaces,
  $\mathcal{H}_\mathrm{cond}$ and $\mathcal{H}_\mathrm{norm}$, such
  that, in the thermodynamic limit,
  $\dim \mathcal{H}_\mathrm{cond}/ \dim \mathcal{H} \to 0$ and the
  free energies of the system restricted to these subspaces cross each
  other for some value of the Hamiltonian parameters, then, the system
  undergoes a first-order quantum phase transition driven by those
  parameters.  The proof is based on an exact probabilistic
  representation of quantum dynamics at an imaginary time identified
  with the inverse temperature of the system. We also show that the
  critical surface has universal features at high and low
  temperatures.
\end{abstract}



\section{Introduction}
The expression ``quantum phase transitions''
(QPTs)~\cite{SGCS,KB,Vojta,Sachdev,Carr,Plastino} usually refers to
phase transitions occurring at zero temperature ($T=0$): in contrast
to classical phase transitions, which are driven by the temperature,
QPTs are meant to be driven by varying some Hamiltonian parameter of
the system.  Such a definition, however, might be a bit misleading. In
more precise terms, one should say that a QPT is characterized by the
existence of a singularity taking place at $T=0$ in correspondence of
some value of the Hamiltonian parameters, the quantum critical point
(QCP).  Here, the quantum nature of the singularity is implicit in the
$T=0$ limit, however, this more precise definition of QPT does not
prevent the phase transition, still purely quantum, to exist also for
$T>0$ via the emergence of a critical line separating the two involved
phases.  Actually, $T=0$ represents just an ideal limit and
understanding the finite temperature counterpart of any QPT is
crucially important.  However, such a task, in general, represents a
quite challenging issue, from both the theoretical and experimental
viewpoints since, above zero temperature, quantum and thermal
fluctuations compete in an intricate manner.

In this work, we present a comprehensive and rigorous approach to a
large class of first-order quantum phase transitions at finite
temperature. An heuristic derivation of this approach as well as some
relevant applications, have been recently presented in \cite{PRL}
while, in the present work, we mainly focus on the proof of the specif
thermodynamic inequalities at the base of our general method.  In the
resulting phase diagram of these first-order QPTs, the critical line
emerging from the QCP establishes a region at finite $T$, which we
call ``condensed'', where the order parameter remains rigidly
invariant. In other words, within the condensed region, the system
behaves like if it were frozen at $T=0$; thermal fluctuations do not
affect the system.  This feature turns out to be particularly
appealing for applications to quantum computing protocols aimed at
finding the ground state (GS) by using low but finite temperatures,
the GS of the condensed phase being the solution of some combinatorial
problem of interest, possibly hard~\cite{Sebenik,Nishimori,Santoro}.

The class of the first-order QPTs we are dealing with was first
introduced and analyzed at $T=0$ in a previous work, where it was
found that the mechanism of the phase transition consists in a
condensation in the space of states~\cite{QPTA}.  Let us consider a
system described by the Hamiltonian
\begin{eqnarray}
  \label{H}
  {H}=\Gamma {K}+{V},
\end{eqnarray}
where $K$ and $V$ are two noncommuting Hermitian operators, $K$ being
dimensionless, and $\Gamma$ a parameter with energy dimensions.  If we
represent $H$ in the eigenbasis of $V$, it is natural to call $V$
``potential operator'', $K$ ``hopping operator'', and $\Gamma$ hopping
parameter. We stress however that $V$ is completely arbitrary and can
involve any kind of particle-interactions.  We will use $\Gamma$ as
the control parameter of the supposed QPT.  Since phase transitions
occur in the thermodynamic limit (TDL), we need a fair competition
between $K$ and $V$ in this limit. If $H$ describes a system of $N$
particles, we assume that the eigenvalues of $K$ and $V$ both scale
linearly with $N$, whereas $\Gamma=O(1)$.  A relevant family of
models to bear in mind concerns qubits based systems.  For these
systems, the space of states $\mathcal{H}$ can be identified with the
space spanned by the $\dim{\mathcal{H}}=2^N$ spin states indicated by
$\ket{\bm{n}} = \ket{n_1} \ket{n_2} \dots \ket{n_N}$, where
$\ket{n_i}=\ket{\pm}$ is an eigenstate of the Pauli matrix
$\sigma^z_i$ relative to the qubit $i=1,\dots,N$, $N$ being the number
of qubits.  The potential $V$ is a diagonal operator in the states
$\ket{\bm{n}}$, namely,
$V=\sum_{\bm{n}} V_{\bm{n}} \ket{\bm{n}}\bra{\bm{n}}$.  The hopping
operator ${K}$ is chosen as the sum of single-flip operators
$ {K}= - \sum_{i=1}^N \sigma^x_i$.  A paradigmatic example of this
family of systems is provided by the Grover Hamiltonian, which
emulates a benchmark model for quantum
search~\cite{Grover,Farhi.Gutmann,Roland.Cerf,Jorg:2008,Jorg:2010},
where $V_{\bm{n}}=-J N \delta_{\bm{n},\bm{n}_1}$,
with $J=O(1)>0$, and $\bm{n}_1$ represents the target of a
totally unstructured (worst case scenario) search.  In contrast,
structured searches correspond to potentials having a smooth minimum
around the target and, therefore, benefit from the application of
gradient-descent based methods like, e.g., in the Ising model where,
however, the corresponding QPTs are second-order.

In Ref.~\cite{QPTA} we have proven the following general result at
$T=0$.  If we can find a partition of the space of states
$\mathcal{H}$ of the system into two subspaces,
$\mathcal{H} = \mathcal{H}_\mathrm{cond} \oplus
\mathcal{H}_\mathrm{norm}$, such that, in the TDL,
$\dim \mathcal{H}_\mathrm{cond}/\dim \mathcal{H} \to 0$ and the ground
state energies of $H$ restricted to these subspaces cross each other
at a finite value of $\Gamma$, then the system undergoes a first-order
QPT driven by this parameter.  \textit{Condensed} and \textit{normal},
the names attributed to the two subspaces, were motivated by the
vanishing of the dimension of $\mathcal{H}_\mathrm{cond}$ relatively
to that of $\mathcal{H}$ in the TDL with the consequence that the QPT
realizes as a condensation in the space of
states.\footnote{\label{NoteNew} It is worth to emphasize
      that, although the condensation that we describe is reminiscent
      of the Bose-Einstein condensation, the latter is a condensation
      in the momentum space, which applies to ideal gases and in which
      the space of states is made by the eigenstates of a
      single-particle Hamiltonian.  On the other hand, we deal with
      interacting particles and our space of states is made by
      many-particle states}.  Condensation QPTs seem
ubiquitous. Besides qubits based systems, they emerge also in
fermionic systems: as we have recently shown, the renowned Wigner
crystallization belongs to this class of QPTs~\cite{WC_QPT}.

As mentioned, the extension of these quantum condensations to finite
temperature has already been presented in the Ref.~\cite{PRL}, where
the phase diagrams at finite $T$ were obtained by simply replacing the
crossing between the ground state energies restricted to the
\textit{condensed} and \textit{normal} subspaces, with the crossing of
the corresponding restricted free energies.  However, while this
generalization at finite $T$ sounds completely natural and physically
appealing, it remains an heuristic argument.  It is the aim of the
present work to provide a rigorous proof.  Whereas the proof for the
$T=0$ case was obtained via an algebraic-functional
approach~\cite{QPTA}, the present proof for the finite temperature
case, which includes the $T=0$ as special limit, is based on an exact
probabilistic representation (EPR) of the quantum dynamics on lattices
introduced some years ago~\cite{BPDAJL}. In contrast to the
algebraic-functional approach, the probabilistic representation of the
quantum dynamics has also the advantage to provide a clear physical
picture. As we will see in detail, linking the imaginary time to the
inverse temperature, we are able to analyze the Gibbs equilibrium at
finite temperature by following trajectories of the system that evolve
for a corresponding finite time and see how the condensations in the
space of states are a consequence of the different crossing rates that
exist for traversing the cond/norm boundary in the two directions, one
being extensive in the system size, $O(N)$, the other being $o(N)$.
In this work, we also show that the critical surface has universal
features at high and low temperatures: in the former case it becomes
proportional to the potential coupling, while in the latter case it
acquires an infinite slope at the QCP.
  
The paper is organized as follows.  Sections II-IV are devoted to the
formal definition of the \textit{condensed} and \textit{normal}
subspaces and to a summary of the $T=0$ formulas, while the main
result at finite $T$ and its proof are provided in Secs. V and VIII,
respectively, Secs. VI and VII being devoted, in this order, to a
proof of the above mentioned universal features and to an application
to the Grover model as an exactly solvable example.  Finally, in
Sec. IX we discuss the equivalence between the
$\dim \mathcal{H}_\mathrm{cond}/ \dim \mathcal{H} \to 0$ condition
with the above mentioned boundary crossing-rate difference.

\section{Normal and condensed subspaces}%
We start by defining a proper partition of the space of states.
Consider a system with Hamiltonian~(\ref{H}), and let
$\{ \ket{\bm{n}_k} \}_{k=1}^{\dim{\mathcal{H}}}$ be a complete
orthonormal set of eigenstates of $V$, \textit{the configurations}:
$V \ket{\bm{n}_k} =V_k \ket{\bm{n}_k}$, $k=1,\dots,\dim{\mathcal{H}}$.
We assume ordered potential values
$V_1 \leq \dots \leq V_{\dim{\mathcal{H}}}$.  For a system of $N$
qubits, for instance, the set of the configurations may correspond to
the set of $\dim{\mathcal{H}}=2^N$ product states of $N$ spins along
some direction, as stated above. For other physical systems composed
by $N_{p}$ particles moving in a lattice of $N$ sites, the filling
$N_{p}/N$ will be assumed to be fixed in the TDL, and the set of the
configurations correspond to all possible ways to accommodate the
$N_{p}$ particles in the $N$ sites according to the fermionic or
bosonic nature of the particles.  In other words, a configuration
$\ket{\bm{n}_k}$ represents the collective positions of the $N_{p}$
indistinguishable particles in the case of fermions and bosons, or the
collective orientations of the $N_{p}=N$ distinguishable qubits
thought fixed at $N$ different spatial positions. At any rate, there
is no limitation in the definition of the set of the configurations
defining the space of states as
$\mathcal{H} = \Span \{ \ket{\bm{n}_k} \}_{k=1}^{\dim{\mathcal{H}}}$
and our general result applies in each abstract or physical case,
without the need of considering separately the nature of the particles
involved, as done for other phase transitions~\cite{AABEB}.
  
Let $\dim{\mathcal{H}}_\mathrm{cond}$ be an integer with
$1\leq \dim{\mathcal{H}}_\mathrm{cond} <\dim{\mathcal{H}}$ and let us
consider a partition of the set of the configurations as
$\{ \ket{\bm{n}_k} \}_{k=1}^{\dim{\mathcal{H}}}=\{
\ket{\bm{n}_k}\}_{k=1}^{\dim{\mathcal{H}}_\mathrm{cond}}\cup \{
\ket{\bm{n}_k}\}_{k=\dim{\mathcal{H}}_\mathrm{cond}+1}^{\dim{\mathcal{H}}}$.
In the Hilbert space of the system,
$\mathcal{H} = \Span \{ \ket{\bm{n}_k} \}_{k=1}^{\dim{\mathcal{H}}}$,
which is equipped with standard complex scalar product $\scp{u}{v}$,
the above partition induces a decomposition of $\mathcal{H}$ as the
direct sum of two mutually orthogonal subspaces, denoted condensed and
normal:
\begin{align}
  \label{DECOMP}
  & \mathcal{H}=\mathcal{H}_\mathrm{cond} \oplus \mathcal{H}_\mathrm{norm},
\end{align}
where
\begin{align}
  \label{DECOMP1}
  & \mathcal{H}_\mathrm{cond} =
    \Span\{\ket{\bm{n}_k}\}_{k=1}^{\dim{\mathcal{H}}_\mathrm{cond}}, \\
  & \mathcal{H}_\mathrm{norm} =
    \Span \{\ket{\bm{n}_k}\}_{k=
    \dim{\mathcal{H}}_\mathrm{cond}+1}^{\dim{\mathcal{H}}}
    =\mathcal{H}_\mathrm{cond}^\perp.
\end{align}
Correspondingly, we define
\begin{align}
  \label{GSE}
  & E=\inf_{\ket{u}\in\mathcal{H}} \frac{\braket{u}{H}{u}}{\scp{u}{u}}, \\
  \label{GSECOND}
  & E_\mathrm{cond}=\inf_{\ket{u}\in\mathcal{H}_\mathrm{cond}}
    \frac{\braket{u}{H}{u}}{\scp{u}{u}}, \\
  \label{GSENORM}
  & E_\mathrm{norm}=\inf_{\ket{u}\in\mathcal{H}_\mathrm{norm}}
    \frac{\braket{u}{H}{u}}{\scp{u}{u}},
\end{align}
which are the GS eigenvalues, respectively, of $H$ and of $H$
restricted to the condensed and normal subspaces.  According to the
scaling properties assumed for $K$ and $V$, we have that $E$,
$E_\mathrm{cond}$ and $E_\mathrm{norm}$ increase linearly with $N$ (at
least in the TDL).

\section{Quantum phase transitions at $T=0$}%
The Hilbert space dimension $\dim{\mathcal{H}}$ generally diverges
exponentially with $N$, while the dimension
$\dim{\mathcal{H}}_\mathrm{cond}$, may or may not be a growing
function of $N$.  In~\cite{QPTA} we have shown that:
\begin{align}
  \label{CONDITION}
  &\mbox{if}\quad \quad \quad
    \lim_{N \to\infty }
    \frac{\dim{\mathcal{H}}_\mathrm{cond}}{\dim{\mathcal{H}}}=0,
  \\
  \label{QPT}
  &\mbox{then} \qquad \lim_{N \to\infty }\frac{E}{N}=
    \lim_{N \to\infty }\min\left\{\frac{E_\mathrm{cond}}{N},
    \frac{E_\mathrm{norm}}{N}\right\}.
\end{align}
For finite sizes, up to corrections $O(1)$, Eq.~(\ref{QPT}) reads
\begin{align}
  \label{QPTR}
  E \simeq \left\{
  \begin{array}{ll}
    E_\mathrm{cond}, \qquad &\mbox{if } E_\mathrm{cond} <E_\mathrm{norm}, 
    \\ 
    E_\mathrm{norm}, \qquad &\mbox{if } E_\mathrm{norm}<E_\mathrm{cond}.
  \end{array}\right.
\end{align}

As a consequence of Eq.~(\ref{QPT}), by varying one or more parameters
of the Hamiltonian $H$, we obtain a QPT, necessarily of first order,
whenever a crossing takes place between $E_\mathrm{cond}$ and
$E_\mathrm{norm}$. In the TDL, the space of states splits at the QCP
(or, more generally, at the quantum critical surface) defined by
\begin{align}
  \label{QPT0}
  \lim_{N \to\infty }\frac{E_\mathrm{cond}}{N}=
  \lim_{N \to\infty }\frac{E_\mathrm{norm}}{N},
\end{align}
and, in correspondence with Eq.~(\ref{QPTR}), for the GS $|E\rangle$
we have either $|E\rangle\in \mathcal{H}_\mathrm{cond}$ or
$|E\rangle\in \mathcal{H}_\mathrm{norm}$.

In order to have a QPT, apart from the necessary condition
(\ref{CONDITION}), $\dim{\mathcal{H}}_\mathrm{cond}$ should also be
chosen in such a way that Eq.~(\ref{QPT0}) admits a finite
solution~\cite{WC_QPT}.  As a general criteria,
$\mathcal{H}_\mathrm{cond}$ should be not too small and not too large
so that neither of the two restrictions of $H$, to
$\mathcal{H}_\mathrm{cond}$ and to $\mathcal{H}_\mathrm{norm}$, have a
QPT. In other words, we want that, in the TDL, $E_\mathrm{cond}/N$ and
$E_\mathrm{norm}/N$ remain both analytic functions of the Hamiltonian
parameters, whereas $E/N$ becomes non-analytic at the
QCP~\cite{WC_QPT}.

\section{Order parameter at $T=0$}
The interpretation of the above class of QPTs in terms of a
condensation in the space of states holds in general, even when
$\mathcal{H}_\mathrm{cond}$ contains many eigenstates of
$V$~\cite{WC_QPT}.  At zero temperature, the probability for the
condensed subspace to be occupied is
\begin{align}
  \label{pcond1}
  p_\mathrm{cond} = \sum_{\ket{\bm{n}_k}\in\mathcal{H}_\mathrm{cond}}
  \abs{\scp{\bm{n}_k}{E}}^2.
\end{align}
On the other hand, in the TDL, since it is either
$\ket{E}\in \mathcal{H}_\mathrm{cond}$ or
$\ket{E}\in \mathcal{H}_\mathrm{norm}$, we find either $p=1$ or $p=0$,
respectively (we assume $\ket{E}$ normalized).  In other words,
$p_\mathrm{cond}$ represents an order parameter of these first-order
QPTs.

We stress that condensation QPTs are intrinsically first-order, for
they can be driven by using even one single Hamiltonian parameter.  In
contrast, as for the classical case, jumps of the order parameter can
result when crossing the coexistence line of two different phases that
originate from the critical point of a second-order QPT.  Notice that,
for such a scenario to take place at zero temperature, one needs that
the Hamiltonian depends on at least two independent parameters (think
to the 1d Ising model in the presence of both a transverse and a
longitudinal magnetic field~\cite{Continentino,Pelissetto}).

\section{Finite temperature quantum condensations}
Our aim is to extend the above class of condensation QPTs to finite
temperature. We suppose that the system, in contact with a heat bath,
is at canonical equilibrium at temperature $T=1/(k_B \beta)$, i.e., it
is in the state described by the Gibbs density matrix operator
$\rho=e^{-\beta H}/\tr e^{-\beta H}$.

Analogously to the $T=0$ case, we proceed by defining the Gibbs free
energies associated to the spaces
$\mathcal{H}, \mathcal{H}_\mathrm{cond}, \mathcal{H}_\mathrm{norm}$,
\begin{align}
  \label{F}
  &e^{-\beta F}
    =
    \tr e^{-\beta H}
    =
    \sum_{\ket{\bm{n}}\in\mathcal{H}}
    \braket{\bm{n}}{e^{-\beta H}}{\bm{n}},
  \\
  \label{Fcond}
  & e^{-\beta F_\mathrm{cond}} 
    =
    \trcond e^{-\beta H_\mathrm{cond}} 
    =
    \sum_{\ket{\bm{n}}\in\mathcal{H}_\mathrm{cond}}
    \braket{\bm{n}}{e^{-\beta H_\mathrm{cond}}}{\bm{n}},
  \\
  \label{Fnorm}
  & e^{-\beta F_\mathrm{norm}}
    =
    \trnorm e^{-\beta H_\mathrm{norm}} 
    =
    \sum_{\ket{\bm{n}}\in\mathcal{H}_\mathrm{norm}}
    \braket{\bm{n}}{e^{-\beta H_\mathrm{norm}}}{\bm{n}},
\end{align}
where $H_\mathrm{cond}$ and $H_\mathrm{norm}$ are the restrictions of
$H$ to the condensed and normal subspaces.\footnote{\label{Note0} In
  the representation of the eigenstates of $V$, $H_\mathrm{cond}$
  corresponds to a null matrix except for the block
  $\braket{\bm{n}_k}{H_\mathrm{cond}}{\bm{n}_{k'}} =
  \braket{\bm{n}_k}{H}{\bm{n}_{k'}}$,
  $k,k'=1,\dots,M_\mathrm{cond}$. Analogously, $H_\mathrm{norm}$
  corresponds to a null matrix except for the block
  $\braket{\bm{n}_{k}}{H_\mathrm{norm}}{\bm{n}_{k'}} =
  \braket{\bm{n}_{k}}{H}{\bm{n}_{k'}}$,
  $k,k'=M_\mathrm{cond}+1,\dots,M$.} Note that
$H_\mathrm{cond}+ H_\mathrm{norm} \neq H$.  It is natural to
investigate whether Eq.~(\ref{QPT}) can be generalized to finite
temperature just by substituting the energies
$E,E_\mathrm{cond},E_\mathrm{norm}$ with the free energies
$F,F_\mathrm{cond},F_\mathrm{norm}$, which scale linearly with $N$
too.

For any partition
$\mathcal{H}=\mathcal{H}_\mathrm{cond} \oplus
\mathcal{H}_\mathrm{norm}$, we will prove that ($X$ stands for either
$\mathrm{cond}$ or $\mathrm{norm}$ and $Y$ for its complement)
\begin{align}
  \label{MATRIX}
  &1 \leq \frac{\braket{\bm{n}}{e^{-\beta H}}{\bm{n}}}
    {\braket{\bm{n}}{e^{-\beta H_{X}}}{\bm{n}}}
    \leq
    e^{\beta \Gamma \min\{A_{X}^{(\mathrm{out})},
    A_{Y}^{(\mathrm{out})}\}},
    \qquad \ket{\bm{n}}\in\mathcal{H}_{X},\\
  \label{upper}
  & F\leq \min\{F_\mathrm{cond},F_\mathrm{norm}\}, \\
  \label{lower}
  & F\geq \min\{F_\mathrm{cond},F_\mathrm{norm}\}
    -\min\{A_{\mathrm{cond}}^{(\mathrm{out})},
    A_{\mathrm{norm}}^{(\mathrm{out})}\}\Gamma, 
\end{align}
where
$A_{X}^{(\mathrm{out})} = \sup_{\ket{\bm{n}}\in\mathcal{H}_{X}}
\sum_{\ket{\bm{n}'}\in\mathcal{H}_\mathrm{Y}}
|\braket{\bm{n}}{K}{\bm{n}'}|$ represents the maximum number of
outgoing links (nonzero matrix elements of $K$) from $\mathcal{H}_{X}$
to $\mathcal{H}_{Y}$. The product
$\min\{A_{X}^{(\mathrm{out})},A_{Y}^{(\mathrm{out})}\}\Gamma$
determines approximately the rate of convergence to 1 of the
probability for crossing the boundary between $\mathcal{H}_{X}$ and
$\mathcal{H}_{Y}$ along the quantum dynamics at imaginary times (see
Sec. VIII).  In the Grover model, e.g.,
$A_{\mathrm{norm}}^{(\mathrm{out})}=1$ while
$A_{\mathrm{cond}}^{(\mathrm{out})}=N$.  As we show in Sec. IX, the
important point is that, in most of the systems of interest, the
conditions $\dim{\mathcal{H}}_{\mathrm{cond}}/\dim{\mathcal{H}}\to 0$
and $A_{\mathrm{norm}}^{(\mathrm{out})}/N\to 0$ are equivalent and,
under any of these conditions, Eqs.~(\ref{upper}) and (\ref{lower}),
up to a term $o(N)$, provide the natural generalization of
Eq.~(\ref{QPTR})
\begin{align}
  \label{TH_QPTR}
  F \simeq \left\{
  \begin{array}{ll}
    F_\mathrm{cond}, \qquad &\mbox{if } F_\mathrm{cond}<F_\mathrm{norm}, 
    \\ 
    F_\mathrm{norm}, \qquad &\mbox{if } F_\mathrm{norm}<F_\mathrm{cond}.
  \end{array} \right.
\end{align}
Equation~(\ref{TH_QPTR}) extends the $T=0$ QPT to finite temperature.
The crossing between $F_\mathrm{cond}$ and $F_\mathrm{norm}$ gives
rise to a first order phase transition controlled by Hamiltonian
parameters and temperature, the equation for the critical surface
being
\begin{align}
  \label{TH_QPTcr}
  \lim_{N \to\infty }\frac{F_\mathrm{cond}}{N}=
  \lim_{N \to\infty }\frac{F_\mathrm{norm}}{N}.
\end{align}
Hereafter, we assume
$\min\{A_{\mathrm{cond}}^{(\mathrm{out})},
A_{\mathrm{norm}}^{(\mathrm{out})}\}=A_{\mathrm{norm}}^{(\mathrm{out})}$.

The probability for the condensed subspace to be occupied represents
an order parameter also at finite temperature and the phase transition
can be interpreted as a condensation in the space of states. In fact,
due to Eqs.~(\ref{MATRIX})
\begin{align}
  \label{pcond3}
  p_\mathrm{cond}= 
  \sum_{\ket{\bm{n}}\in\mathcal{H}_\mathrm{cond}}
  \braket{\bm{n}}{\rho}{\bm{n}}
  \simeq 
  \frac{1}{1+e^{-\beta(F_\mathrm{norm}-F_\mathrm{cond})}},
\end{align}
where the equality holds in the TDL with $p_\mathrm{cond}=1$ in the
condensed phase $F_\mathrm{cond}<F_\mathrm{norm}$ and
$p_\mathrm{cond}=0$ in the normal one
$F_\mathrm{norm}<F_\mathrm{cond}$.  At the critical surface separating
the two phases we have $p_\mathrm{cond}=1/2$.

Equations~(\ref{upper})-(\ref{lower}) are easily derived from
Eqs.~(\ref{MATRIX}).  Before giving the proof of Eqs.~(\ref{MATRIX}),
we illustrate some universal features of the finite temperature
condensations and the application of our findings to the Grover model.

\section{Universal features of the critical surface} We recall that
standard canonical thermodynamics relations such as $F=U-TS$, $U$
being the internal energy and $S=-\partial F/\partial T$ the entropy,
apply also to the quantum case.  Suppose that the potential $V$ in the
Hamiltonian~(\ref{H}) depends on a single parameter, say $J$, having
energy dimensions: $V=J\tilde{V}$, $\tilde{V}$ being dimensionless.
In this case, keeping fixed the kinetic parameter $\Gamma$, the
equation for the critical surface~(\ref{TH_QPTcr}) determines the
critical temperature as a function of $J$: $T=T(J)$.  The critical
temperature $T(J)$ is the $N\to\infty$ limit of the ``finite size
critical temperature'' $T_N(J)$ determined by the finite size
analogous of Eq.~(\ref{TH_QPTcr}).  Under the mild assumption that
$T_N(J)$ converges uniformly to $T(J)$ we are allowed to exchange the
order of limits $N\to \infty$ with $J\to \infty$ and also to exchange
the order of the limit $N\to \infty$ with the derivative $d/dJ$.  In
the following, we shall make use of these properties to establish two
universal features of the critical temperature: at large potential
values, $J\to\infty$, and at the QCP, $J\to J_c$. In both cases the
starting point is Eq.~(\ref{TH_QPTcr}) at finite size rewritten as
\begin{align}
  \label{UNIV0}
  T_N(J)= \frac{U_\mathrm{norm}-U_\mathrm{cond}}
  {S_\mathrm{norm}-S_\mathrm{cond}}.
\end{align}

Let us consider the limit $J\to \infty$. Here, all the eigenvalues of
$H$, as well as all the eigenvalues of the restrictions of $H$ to the
normal and condensed subspaces, become proportional to $J$. As a
consequence, the internal energies $U_\mathrm{norm}$ and
$U_\mathrm{cond}$ become also proportional to $J$. It follows that
$U_\mathrm{norm}-U_\mathrm{cond}=\alpha N J$, where $\alpha$ is a
constant independent of $J$.  On the other hand, for the entropy (of
the whole space and, similarly, of the restrictions), we have
\begin{align}
  \label{ENTROPY}
  S=k_\mathrm{B} \log\left(\tr e^{- \beta_N(J)H}\right)+\frac{U}{T_N(J)}.
\end{align}
From Eq.~(\ref{ENTROPY}) we see that, by assuming $T_N(J)=\gamma J$,
where $\gamma$ is a finite positive constant, in the limit
$J\to\infty$, the entropy (of the whole space and, similarly, of the
restrictions) becomes independent of $J$.  On combining this fact with
Eq.~(\ref{UNIV0}) we see that, in the limit $J\to \infty$,
$T_N(J)=\gamma J$ is solution of Eq.~(\ref{UNIV0}).  Finally, in the
TDL, taking into account that $S_\mathrm{norm}$ must be extensive
($S_\mathrm{cond}$ could be extensive or not), we get the value of
$\gamma$ as follows
\begin{align}
  \label{UNIV01}
  \gamma=\lim_{J\to\infty}\frac{T(J)}{J}=
  \lim_{J\to\infty}\lim_{N\to\infty}
  \frac{U_\mathrm{norm}-U_\mathrm{cond}}{J(S_\mathrm{norm}-S_\mathrm{cond})}.
\end{align}   

Let us now consider the limit $J\to J_c$.  By using
$\partial F/\partial T=-S$, and, similarly, for the restrictions to
the subspaces, we can evaluate the total derivative of
Eq.~(\ref{TH_QPTcr}) with respect to $J$ as
\begin{align}
  \label{UNIV}
  \frac{\partial(F_\mathrm{norm}-F_\mathrm{cond})}{\partial J}
  -(S_\mathrm{norm}-S_\mathrm{cond})\frac{\partial T_N(J)}{\partial J}=0,
\end{align}   
which provides
\begin{align}
  \label{UNIV1}
  \frac{\partial T_N(J)}{\partial J}=
  \frac{\partial(F_\mathrm{norm}-F_\mathrm{cond})}{\partial J}
  \frac{1}{S_\mathrm{norm}-S_\mathrm{cond}}.
\end{align}
Again we observe that the free energies and, at any finite $T$, also
the entropies, are extensive quantities. This implies that, for any
finite $T$, the TDL of Eq.~(\ref{UNIV1}) is finite.  However, since
$J\to J_c$ implies $T\to 0$, disregarding cases like spin-glass
models, the entropy density of the system as well as of its
restrictions tend to zero in the TDL.  Let us assume that, for
$J=J_c$, in the TDL we have
${\partial(E_\mathrm{norm}/N-E_\mathrm{cond}/N)}/{\partial J}\neq 0$.
From Eq.~(\ref{UNIV1}) we conclude that
\begin{align}
  \label{UNIV2}
  \lim_{J\to J_c}\frac{\partial T(J)}{\partial J}
  =\lim_{J\to J_c}\lim_{N\to\infty}\frac{\partial T_N(J)}{\partial J}
  =+\infty.
\end{align}
We can show that Eq.~(\ref{UNIV2}) holds true also when, for $J=J_c$,
in the TDL we have
${\partial(E_\mathrm{norm}/N-E_\mathrm{cond}/N)}/{\partial J}= 0$ but
${\partial^2(E_\mathrm{norm}/N-E_\mathrm{cond}/N)}/{\partial J^2}\neq
0$ and possibly infinite.  The argument is based on the assumption
that, in the limit $T\to 0$, the specific heat of the two restrictions
tend to 0 faster than $T$, or else that
${\partial^2(E_\mathrm{norm}/N-E_\mathrm{cond}/N)}/{\partial J^2}\to
\infty$ in the TDL, as indeed occurs in many cases of interest.

Clearly, what actually matters is the ratio $J/\Gamma$. In fact, by
using the same arguments we can equivalently rewrite
Eq.~(\ref{UNIV01}) as $\gamma=\lim_{J/\Gamma\to\infty} T(J)/J$ or
else, if the potential parameter $J$ is kept constant and $\Gamma$ is
varied, as $\gamma=\lim_{\Gamma/J\to 0} T(\Gamma)/J$ and, similarly,
we can rewrite Eq.~(\ref{UNIV2}) as
$\lim_{\Gamma\to \Gamma_c} \partial T(\Gamma)/\partial \Gamma
=+\infty$, where $T(\Gamma)$ is the critical temperature in the limit
$N\to\infty$ and $\Gamma_c$ provides the QCP. As we shall see in the
next Section, the constant $\gamma$ can be easily evaluated in the
exactly solvable Grover model where $\gamma=1/(k_\mathrm{B} \log 2)$.

\section{The Grover model as an exactly solvable paradigmatic example}

\subsection{The case $T=0$}
Here $V_1=-JN$, with $J>0$, and $V_k=0$, for
$k=2,3,\dots,\dim{\mathcal{H}}=2^N$. We can assume
$\dim{\mathcal{H}}_\mathrm{cond}=1$ independent of $N$. We find
$\ket{E_\mathrm{cond}}=\ket{\bm{n}_1}$ and $E_\mathrm{cond}=V_1$. Up
to a correction $\BigO{N/\dim{\mathcal{H}}}$, we also have
$E_\mathrm{norm}=-\Gamma N$~\cite{QPTA}. Therefore Eq.~(\ref{QPTR})
becomes
\begin{align}
  \label{QPTR1}
  E\simeq
  \left\{
  \begin{array}{ll}
    -JN, \qquad &\mbox{if } \Gamma<\Gamma_\mathrm{c},
    \\ 
    -\Gamma N, \qquad &\mbox{if } \Gamma>\Gamma_\mathrm{c},
  \end{array}\right.
\end{align}   
where the QCP, $\Gamma_\mathrm{c}$, is determined by Eq.~(\ref{QPT0}),
namely, $\Gamma_\mathrm{c}=J$.  For $\Gamma>\Gamma_\mathrm{c}$ the GS
of the model coincides with the GS of the hopping operator $K$, while
for $\Gamma<\Gamma_\mathrm{c}$ the system stays locked in the
configuration $\ket{\bm{n}_1}$.  We thus have a QPT that corresponds
to a condensation in the space of states.

\subsection{The general case $T\geq 0$}
\begin{figure}
  \begin{center}
    {\includegraphics[width=0.63\columnwidth,clip]{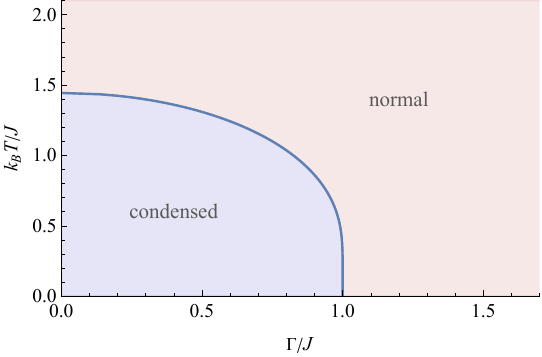}}
    \caption{ Phase diagram for the Grover model at thermal
      equilibrium, the solid line separating the two phases is drawn
      according to Eq.~(\ref{GammacT_Grover}).}
    \label{phase_diagram_Grover}
  \end{center}
\end{figure}

Since $\dim{\mathcal{H}}_\mathrm{cond}=1$, we have
$-\beta F_\mathrm{cond}= -\beta V_1$ with $V_1=-JN$.  Up to
corrections exponentially small in $N$, the free energy of the normal
subspace coincides with that of the hopping operator $K$ whose levels
are $-\Gamma(N-2j)$, $j=0,\dots,N$, and have degeneracy
$N!/(j!(N-j)!)$,
\begin{align}
  e^{-\beta F_\mathrm{norm}} = \tr e^{-\beta K}
  = \sum_{j=0}^{N} \binom{N}{j} e^{-\beta(-\Gamma(N-2j))},
\end{align}
which provides
$ -\beta F_\mathrm{norm} =N\log \left( 2\cosh(\beta\Gamma) \right)$.
The critical surface defined by Eq.~(\ref{TH_QPTcr}) is thus given by
$ \log \left( 2\cosh(\beta\Gamma)\right) = \beta J$ (a result also
found in Ref.~\cite{Jorg:2010} via approximate methods) which can be
solved to explicitly provide
$\Gamma_\mathrm{c}=\Gamma_\mathrm{c}(T)$\footnote{\label{NoteExplicit}On
  posing $x=e^{\beta \Gamma}$, we are left with the reciprocal
  equation $\log(x+x^{-1})=\beta J$ that can, in turn, be transformed
  into a quadratic equation for $x$ having two real and positive
  roots: one with $x>1$, which corresponds to a positive $\Gamma$
  (Eq.~\ref{GammacT_Grover}), and the other one with $x<1$, which
  corresponds to a negative $\Gamma$.},
\begin{align}
  \label{GammacT_Grover}
  \Gamma_\mathrm{c}(T) =
  J + k_B T \log \left(
  \frac{1}{2} +\sqrt{\frac{1}{4}- e^{-2J/(k_BT)}} \right).
\end{align}
Note that Eq.~(\ref{GammacT_Grover}) is defined only for
$k_BT\leq J/\log 2$ and for $T\to 0^+$ returns the already analyzed
$T=0$ QPT.  A parametric plot of $(\Gamma_\mathrm{c}(T),T)$ is shown
in Fig.~\ref{phase_diagram_Grover}.  The shaded area is the condensed
phase.  No condensed phase is possible for
$\Gamma>\Gamma_\mathrm{c}(0)=J$ (point of minimal entropy).  For
$0\leq \Gamma \leq \Gamma_\mathrm{c}(0)$ the condensed phase extends
to the finite temperature $T_\mathrm{c}(\Gamma)$ obtained inverting
Eq.~(\ref{GammacT_Grover}). No condensed phase is possible for
$T>T_\mathrm{c}(0)=J/(k_B\log 2)$ (point of maximal entropy).

Thermodynamics follows easily: internal energies $U_\mathrm{cond}=-JN$
and $U_\mathrm{norm}=-\Gamma N \tanh(\beta \Gamma)$; entropies
$S_\mathrm{cond}=0$ and
$S_\mathrm{norm}=N k_B [\log
(2\cosh(\beta\Gamma))-\beta\Gamma\tanh(\beta \Gamma)]$; specific heats
$c_\mathrm{cond}=0$ and
$c_\mathrm{norm}=k_B (\beta\Gamma/\cosh(\beta \Gamma))^2$.  Notice
that, whereas the free energy $F$ is always continuous in $T$, the
internal energy $U$, the entropy $S$, and the specific heat $c$, are
all discontinuous along any curve that crosses the critical surface,
except for $T\to 0$.  This in particular implies a non null latent
heat proportional to the entropy of the normal phase:
$U_\mathrm{norm}-U_\mathrm{cond}|_{T=T_c}=k_B T_c
S_\mathrm{norm}|_{T=T_c}$.  This latent heat represents the minimal
amount of energy to be subtracted from the system in order to bring it
from the normal to the condensed phase.

\section{Proof of Eqs.~(\ref{MATRIX})}%
In the following, we prove the lower and upper bounds of
Eq.~(\ref{MATRIX}).  The starting point is the exact probabilistic
representation of the quantum evolution introduced
in~\cite{BPDAJL}. According to this EPR, at imaginary time $t$, to be
identified here with the inverse temperature $\beta$, we have
($\hbar=1$)
\begin{align}
  \label{EPR0}
  \braket{\bm{n}}{e^{-Ht}}{\bm{n}_{0}} =  
  \E  \left(  
  \mathcal{M}^{[0,t)}_{\bm{n}_{0}} \delta_{ \bm{n}_{N_t} , \bm{n} }
  \right), 
\end{align}
where $\E(\cdot)$ is the probabilistic expectation over the continuous
time Markov chain of configurations
$\bm{n}_{0},\bm{n}_{s_1},\dots,\bm{n}_{s_{N_t}}$ (hereafter, named
trajectory) defined by the transition matrix
\begin{align}
  \label{Pf} 
  P_{\bm{n},\bm{n}'} = \frac{|\braket{\bm{n}}{K}{\bm{n}'}|}{A(\bm{n})},
  \qquad A(\bm{n}) = \sum_{\bm{n}'} |\braket{\bm{n}}{K}{\bm{n}'}|,
\end{align}
and the sequence of jumping times $s_1,s_2,\dots,s_{N_t}$ obtained
from the Poissonian conditional probability density
\begin{align}
  P(s_{k}|s_{k-1}) =
  e^{-\Gamma A(\bm{n}_{s_{k-1}}) (s_{k}-s_{k-1})}
  \Gamma A(\bm{n}_{s_{k-1}}),
  \label{Ps}
\end{align}
$N_t$ being the number of jumps occurred along the trajectory before
the time $t$.  Note that hereafter with the term
\textit{configuration} we may indicate the eigenstate of $V$,
$\ket{\bm{n}}$, or the set of indices $\bm{n}$ which define such
state. The integer $A(\bm{n})$ is called the number of links, or
degree, of $\bm{n}$ and represents the number of non null off-diagonal
matrix elements $\langle \bm{n} |H|\bm{n}' \rangle$.  Starting form
the configuration $\bm{n}_{0}$ at time $s_0=0$, we draw a
configuration $\bm{n}_{s_1}$ with probability
$P_{\bm{n}_{0},\bm{n}_{s_1}}$ at time $s_1$ drawn with probability
density $P(s_{1}|s_{0})$, then we draw a configuration $\bm{n}_{s_2}$
with probability $P_{\bm{n}_{s_1},\bm{n}_{s_2}}$ at time $s_2$ drawn
with probability density $P(s_{2}|s_{1})$, and so on until we reach
the configuration $\bm{n}_{N_t}$ at time $s_{N_t}$ such that
$s_{N_t+1}>t$.\footnote{\label{Note1}Note that the Poisson processes
  associated to each jump are defined left continuous~\cite{BPDAJL},
  as a consequence, the configuration $\bm{n}_{N_t}$ is the one
  realized by the Markov chain just before the final time $t$.}  The
stochastic functional $\mathcal{M}^{[0,t)}_{\bm{n}_0}$ is then defined
as
\begin{align}
  \label{EPR}
  \mathcal{M}^{[0,t)}_{\bm{n}_0}
  = e^{\sum_{k=0}^{N_t-1} [\Gamma A(\bm{n}_{s_k})
  -V(\bm{n}_{s_k})](s_{k+1}-s_{k})}
  ~
  e^{[\Gamma A(\bm{n}_{s_{N_t}})
  -V(\bm{n}_{s_{N_t}})](t-s_{N_t})}.
\end{align}
Whereas a more general formulation of the EPR is
possible~\cite{BPDAJL}, that presented above holds in the
statistically manageable case in which no sign problem arises, e.g.,
when $\braket{\bm{n}}{K}{\bm{n}'}\leq 0$ for any $\bm{n},\bm{n}'$.  We
assume to be in this class of ``bosonic'' systems. In particular, for
qubit systems $K$ is the sum of single flip operators, for which
$\braket{\bm{n}}{K}{\bm{n}'}=0,-1$.  If the whole set of
configurations is connected by ${K}$, as we assume, the Markov chain
is ergodic with invariant measure
$\pi_{\bm{n}}= A(\bm{n})/\sum_{\bm{n}'} A(\bm{n}')$.  For example, in
qubit systems as the Grover model, the degree of the configurations is
constant, $A(\bm{n})=N$, and $\pi_{\bm{n}}=1/\dim{\mathcal{H}}$.

Let us indicate by
$\widehat{\mathcal{H}}= \{\bm{n}_k
\}_{k=1}^{\dim{\mathcal{H}}_\mathrm{cond}}$
($\widetilde{\mathcal{H}}=
\{\bm{n}_k\}_{k=\dim{\mathcal{H}}_\mathrm{cond}+1}^{\dim{\mathcal{H}}}$)
the set of configurations defining the states in
$\mathcal{H}_\mathrm{cond}$ ($\mathcal{H}_\mathrm{norm}$).  A generic
configuration of $\widehat{\mathcal{H}}$ ($\widetilde{\mathcal{H}}$)
will be indicated by $\widehat{\bm{n}}$ ($\widetilde{\bm{n}}$).  For
any configuration $\bm{n}=\widehat{\bm{n}}$ or
$\bm{n}=\widetilde{\bm{n}}$ we can always split its degree as
\begin{align}
  \label{Ainout}
  A(\bm{n})=A^{(\mathrm{in})}(\bm{n})
  +A^{(\mathrm{out})}(\bm{n}),
\end{align}
where $A^{(\mathrm{in})}(\bm{n})$ and $A^{(\mathrm{out})}(\bm{n})$
represent the number of links connecting $\bm{n}$ with configurations
inside or outside its membership subset, $\widehat{\mathcal{H}}$ or
$\widetilde{\mathcal{H}}$.\footnote{\label{Note2}Note that
  $A^{(\mathrm{in})}(\widehat{\bm{n}})$ and
  $A^{(\mathrm{in})}(\widetilde{\bm{n}})$ represent the number of non
  null off-diagonal matrix elements
  $\langle \widehat{\bm{n}} |H_{\mathrm{cond}}|\widehat{\bm{n}}'
  \rangle$ and
  $\langle \widetilde{\bm{n}} |H_{\mathrm{norm}}|\widetilde{\bm{n}}'
  \rangle$, respectively.}

Consider trajectories beginning and ending at a configuration
$\widetilde{\bm{n}}$ of $\widetilde{\mathcal{H}}$.  Introducing the
random variable $K_t=0,1,2,\dots$ counting the number of times the
Markov chain transits throughout $\widehat{\mathcal{H}}$ in the
interval $[0,t)$, we decompose the expectation as a sum of two
constrained expectations as follows
\begin{align}
  \label{EPR1}
  \braket{\widetilde{\bm{n}}}{e^{-Ht}}{\widetilde{\bm{n}}}
  = \E\left(
  {\cal M}^{[0,t)}_{\widetilde{\bm{n}}}
  \delta_{ \bm{n}_{N_t} , \widetilde{\bm{n}} };K_t=0 \right)
  +\E\left(
  {\cal M}^{[0,t)}_{\widetilde{\bm{n}}}
  \delta_{ \bm{n}_{N_t} , \widetilde{\bm{n}} };K_t\geq 1 \right).
\end{align}

Consider the term $K_t=0$.  Each trajectory contributing to this event
is characterized by a sequence of $N_t$ jumping times
$s_1,~s_2,\dots,s_{N_t}$ extracted along a sequence of configurations
$\widetilde{\bm{n}},\widetilde{\bm{n}}_1,\widetilde{\bm{n}}_2,\dots,
\widetilde{\bm{n}}_{N_t}$.  Hence, regardless of any other detail,
such a trajectory is realized if none of the associated out links
jump, which occurs with probability
$\exp\{-\Gamma[A^{(\mathrm{out})}(\widetilde{\bm{n}})s_1+
A^{(\mathrm{out})}(\widetilde{\bm{n}}_1)(s_2-s_1)+ \dots
+A^{(\mathrm{out})}(\widetilde{\bm{n}}_{N_t-1})(s_{N_t}-s_{N_t-1})+
A^{(\mathrm{out})}(\widetilde{\bm{n}}_{N_t})(t-s_{N_t})]\}$.  On the
other hand, Eq.~(\ref{EPR}) shows that along the same trajectory the
hopping term provides the weight
$\exp\{\Gamma[A(\widetilde{\bm{n}})s_1+A(\widetilde{\bm{n}}_1)(s_2-s_1)+
\dots +A(\widetilde{\bm{n}}_{N_t-1})(s_{N_t}-s_{N_t-1})+
A(\widetilde{\bm{n}}_{N_t})(t-s_{N_t})]\}$.  By using
$A(\widetilde{\bm{n}})-A^{(\mathrm{out})}(\widetilde{\bm{n}}) =
A^{(\mathrm{in})}(\widetilde{\bm{n}})$, we obtain
\begin{align}
  \label{Kt=0}
  \E\left(
  {\cal M}^{[0,t)}_{\widetilde{\bm{n}}}
  \delta_{ \bm{n}_{N_t} , \widetilde{\bm{n}} };K_t=0 \right)
  &=
    \E\left(
    \widetilde{{\cal M}}^{[0,t)}_{\widetilde{\bm{n}}}
    \delta_{ \bm{n}_{N_t} , \widetilde{\bm{n}} } \right)
    \nonumber \\
  &=
    \braket{\widetilde{\bm{n}}}{e^{-H_\mathrm{norm}t}}{\widetilde{\bm{n}}},
\end{align}
where $\widetilde{{\cal M}}^{[0,t)}_{\widetilde{\bm{n}}}$ is the
stochastic functional defined in terms of
$H_\mathrm{norm}$\footnote{See footnote \ref{Note2}} and
Eq.~(\ref{EPR0}) has been used again (now applied to the system
governed by $H_\mathrm{norm}$) to get the second equality.  Since
${{\cal M}}^{[0,t)}_{\widetilde{\bm{n}}}>0$, Eqs.~(\ref{EPR1}) and
(\ref{Kt=0}) give
\begin{align}
  \label{uppernorm}
  \braket{\widetilde{\bm{n}}}{e^{-Ht}}{\widetilde{\bm{n}}}
  \geq \braket{\widetilde{\bm{n}}}{e^{-H_\mathrm{norm}t}}{\widetilde{\bm{n}}}.
\end{align}
Considering trajectories beginning and ending at a configuration
$\widehat{\bm{n}}$ of $\widehat{\mathcal{H}}$, we get a similar
relation with $\widetilde{\bm{n}} \to \widehat{\bm{n}}$ and
$H_\mathrm{norm} \to H_\mathrm{cond}$. This completes the proof of the
first inequality in Eq.~(\ref{MATRIX}).

Proving the second inequality of (\ref{MATRIX}) requires the analysis
of the term $K_t\geq 1$ in Eq.~(\ref{EPR1}), which is quite more
involved. We have
\begin{align}
  \label{Kt}
  \E\left(  {\cal M}^{[0,t)}_{\widetilde{\bm{n}}}\delta_{ \bm{n}_{N_t} ,
  \widetilde{\bm{n}} };K_t\geq 1 \right)
  =
  \sum_{\xi}{\cal M}^{[0,t)}_{\widetilde{\bm{n}}}(\xi)
  \mathbb{P}_t(\widetilde{\bm{n}} \xrightarrow{\xi}
  \widetilde{\bm{n}}; K_t\geq 1),
\end{align}
where the sum runs over the ``space-time'' trajectories $\xi$, and
$\mathbb{P}_t(\widetilde{\bm{n}} \xrightarrow{\xi} \widetilde{\bm{n}};
K_t\geq 1)$ stands for the probability that, starting from
$\widetilde{\bm{n}}$, $\xi$ ends in $\widetilde{\bm{n}}$ by transiting
throughout $\widehat{\mathcal{H}}$ at least once within the time
$t$. Apart from $K_t$, each $\xi$ has a probability obtained via the
sequence of jumping links and jumping times according to
Eqs.~(\ref{Pf}) and (\ref{Ps}).  If $A(\bm{n})=N$ is constant, which
happens in many qubit systems, the trajectories have no preferential
directions and, therefore, no correlation with the random variable
$K_t$ (in particular, trajectories visiting the same number of
configurations and corresponding jumping times have the same
probability).  In more general systems, due to the condition
(\ref{CONDITION}), the correlations with $K_t$ become negligible in
the TDL.  We then have
\begin{align}
  \label{Ktb} 
  \E\left(  {\cal M}^{[0,t)}_{\widetilde{\bm{n}}}
  \delta_{ \bm{n}_{N_t} , \widetilde{\bm{n}} };K_t\geq 1 \right)
  &\simeq
    \sum_{\xi}{\cal M}^{[0,t)}_{\widetilde{\bm{n}}}(\xi)
    \mathbb{P}_t(\widetilde{\bm{n}}
    \xrightarrow{\xi}
    \widetilde{\bm{n}})
    \mathbb{P}_t(\widetilde{\bm{n}}; K_t\geq 1)
    \nonumber \\
  &=
    \braket{\widetilde{\bm{n}}}{e^{-Ht}}{\widetilde{\bm{n}}}
    \mathbb{P}_t(\widetilde{\bm{n}}; K_t\geq 1),
\end{align}
where $\mathbb{P}_t(\widetilde{\bm{n}}; K_t\geq 1)$ stands for the
total probability that, within the time $t$ and starting from a given
configuration $\widetilde{\bm{n}}$, the system transits through
$\widehat{\mathcal{H}}$ at least once.  It is clear that, given $N$,
$\mathbb{P}_t(\widetilde{\bm{n}}; K_t\geq 1)\to 1$ for
$t\to\infty$. However, we are interested in the other order of limits
and, actually, here $t$ must be kept finite while extrapolating the
TDL.  In fact, we want a bound for
$\mathbb{P}_t(\widetilde{\bm{n}}; K_t\geq 1)$ in the TDL. We have
\begin{align}
  \label{Ktc}
  \mathbb{P}_t(\widetilde{\bm{n}}; K_t\geq 1)
  \leq 1-\mathbb{P}_t(\widetilde{\bm{n}}; K_t=0).
\end{align}
Notice that $\mathbb{P}_t(\widetilde{\bm{n}}; K_t=0)$ represents the
probability to remain in $\widetilde{\mathcal{H}}$ during the time $t$
and it does not coincide with the complement of
$\mathbb{P}_t(\widetilde{\bm{n}}; K_t\geq 1)$.  In fact, by
definition, if $K_t\geq 1$, $K_t$ counts how many times a trajectory
that starts from $\widetilde{\mathcal{H}}$, transits through
$\widehat{\mathcal{H}}$, and eventually goes back to
$\widetilde{\mathcal{H}}$, while the complement of the event
$K_t\geq 1$ contains also all the trajectories that, starting from
$\widetilde{\mathcal{H}}$, transit through $\widehat{\mathcal{H}}$ a
certain number of times but eventually do not terminate in
$\widetilde{\mathcal{H}}$.  Let $\widetilde{\partial}$ be the boundary
set between $\widetilde{\mathcal{H}}$ and $\widehat{\mathcal{H}}$
belonging to $\widetilde{\mathcal{H}}$:
\begin{align}
  \label{boundary}
  \widetilde{\partial}=\left\{\widetilde{\bm{n}}\in
  \widetilde{\mathcal{H}}:~\exists \widehat{\bm{n}}\in \widehat{\mathcal{H}}
  \textrm{ such that }
  \langle \widetilde{\bm{n}}|K|\widehat{\bm{n}}\rangle\neq 0\right\}.
\end{align}
Clearly, $\widetilde{\partial}$ represents the set of configurations
having the smallest probability of remaining in
$\widetilde{\mathcal{H}}$ and such a probability corresponds to the
event where no jump occurs through the outgoing links of these
boundary configurations.  Therefore, according to Eq.~(\ref{Ps}) and
to the definition (\ref{Ainout}) we have
\begin{align}
  \label{Ktd}
  \mathbb{P}_t(\widetilde{\bm{n}}; K_t=0)
  &\geq
    \inf_{\widetilde{\bm{n}}\in \widetilde{\partial}}
    \mathbb{P}_t(\widetilde{\bm{n}}; K_t=0)
    \nonumber \\
  &=
    \inf_{\widetilde{\bm{n}}\in \widetilde{\partial}}
    e^{-\Gamma A^{(\mathrm{out})}(\widetilde{\bm{n}})t}
    \nonumber \\
  &=
    e^{-\sup_{\widetilde{\bm{n}}\in
    \widetilde{\partial}}\Gamma A^{(\mathrm{out})}(\widetilde{\bm{n}})t}.
\end{align}
In conclusion, we have
\begin{align}
  \label{Kte}
  \sup_{\widetilde{\bm{n}}} \mathbb{P}_t(\widetilde{\bm{n}}; K_t\geq 1)
  \leq 1- e^{-\sup_{\widetilde{\bm{n}}\in
  \widetilde{\partial}}\Gamma A^{(\mathrm{out})}(\widetilde{\bm{n}})t}.
\end{align}
Eq.~(\ref{Kte}) shows that the probability we are interested in has an
upper bound that still goes to 1 exponentially in the TDL, but with a
rate that is not extensive in $N$, in fact,
$A^{(\mathrm{out})}(\widetilde{\bm{n}})$ is not extensive in $N$.
Typically, in qubit systems $A^{(\mathrm{out})}(\widetilde{\bm{n}})$
is $O(1)$, but for our aims it could be also $o(N)$, as it occurs in
system of fermions or hard-core bosons.  Combining Eqs.~(\ref{EPR1}),
(\ref{Kt=0}) and (\ref{Ktb}), and then Eq.~(\ref{Kte}), we obtain
\begin{align}
  \braket{\widetilde{\bm{n}}}{e^{-Ht}}{\widetilde{\bm{n}}}
  \leq 
  \braket{\widetilde{\bm{n}}}{e^{-H_\mathrm{norm} t}}{\widetilde{\bm{n}}}
  +
  \braket{\widetilde{\bm{n}}}{e^{-Ht}}{\widetilde{\bm{n}}}
  \left( 1-e^{-\sup_{\widetilde{\bm{n}}\in
  \widetilde{\partial}}\Gamma A^{(\mathrm{out})}(\widetilde{\bm{n}})t}\right),
\end{align}
or
\begin{align}
  \braket{\widetilde{\bm{n}}}{e^{-Ht}}{\widetilde{\bm{n}}}
  \leq 
  \braket{\widetilde{\bm{n}}}{e^{-H_\mathrm{norm} t}}
  {\widetilde{\bm{n}}}e^{\sup_{\widetilde{\bm{n}}\in
  \widetilde{\partial}}\Gamma A^{(\mathrm{out})}(\widetilde{\bm{n}})t}.
\end{align}
Since
$\sup_{\widetilde{\bm{n}}\in \widetilde{\partial}}
A^{(\mathrm{out})}(\widetilde{\bm{n}}) =
A_{\mathrm{norm}}^{(\mathrm{out})}$ and we assumed
$A_{\mathrm{norm}}^{(\mathrm{out})} <
A_{\mathrm{cond}}^{(\mathrm{out})}$, the second inequality in
Eq.~(\ref{MATRIX}) is proven for $X=\mathrm{norm}$.

To prove the second inequality in Eq.~(\ref{MATRIX}) for
$X=\mathrm{cond}$, we have to proceed in a slightly different way.
Notice, in fact, that the analogous of Eq.~(\ref{Kte}) for the set
$\widehat{\mathcal{H}}$ also holds, but it is of little use because,
in general, whereas $A^{(\mathrm{out})}(\widetilde{\bm{n}})$ is not
extensive in $N$, $A^{(\mathrm{out})}(\widehat{\bm{n}})$ can be
extensive in $N$. In fact, this is just the case of the Grover model
previously analyzed, as well as the case of regular qubit systems.
Therefore, we avoid using Eq.~(\ref{Kte}) for $\widehat{\mathcal{H}}$
directly.  The main idea here is that, due to the fact that
$\widehat{\mathcal{H}}$ is a small portion of the whole set of
configurations, the probability for a trajectory starting from
$\widehat{\mathcal{H}}$ to reach $\widetilde{\mathcal{H}}$, approaches
1 exponentially (in both $t$ and $N$) with a large rate, but once it
is in $\widetilde{\mathcal{H}}$, the probability that it goes back to
$\widehat{\mathcal{H}}$ approaches 1 with the same identical small
rate of Eq.~(\ref{Kte}). Let us make concrete this idea by explicitly
taking into account just one jump into $\widetilde{\mathcal{H}}$ as
follows
\begin{align}
  \label{Ktf}
  \sup_{\widehat{\bm{n}}} \mathbb{P}_t(\widehat{\bm{n}}; L_t\geq 1)
  &=
    \sup_{\widehat{\bm{n}}\in\widehat{\partial}}
    \mathbb{P}_t(\widehat{\bm{n}}; L_t\geq 1)
    \nonumber \\
  &\simeq
    \sup_{\widehat{\bm{n}}\in\widehat{\partial}} ~
    \sum_{\widetilde{\bm{n}}\in \mathcal{A}^{(\mathrm{out})}(\widehat{\bm{n}})}
    \int_0^t ds \Gamma e^{-\Gamma A(\widehat{\bm{n}})s}
    \mathbb{P}_t(\widetilde{\bm{n}}; Q_{t-s}\geq 1)
\end{align}
where $L_t$ and $\widehat{\partial}$ are the analogous of $K_t$ and
$\widetilde{\partial}$ for $\widehat{\mathcal{H}}$,
$\mathcal{A}^{(\mathrm{out})}(\widehat{\bm{n}})$ is the set of
configurations in $\widetilde{\mathcal{H}}$ which are first neighbors
of $\widehat{\bm{n}}$ (whose number is
$A^{(\mathrm{out})}(\widehat{\bm{n}})$), $s$ is a random time at which
a jump toward one configuration
$\widetilde{\bm{n}}\in \mathcal{A}^{(\mathrm{out})}(\widehat{\bm{n}})
\subset \widetilde{\partial}$ takes place, and $Q_{t}$ counts the
number of times the trajectory that starts from
$\widetilde{\mathcal{H}}$ leaves $\widetilde{\mathcal{H}}$ by ending
in $\widehat{\mathcal{H}}$ within the time interval $[0,t)$.  Equation
(\ref{Ktf}) holds approximately because we have neglected the
trajectories that, starting from $\widetilde{\partial}$, reach for the
first time $\widehat{\mathcal{H}}$ by using more than one jump
within the time interval $[0,t)$. However, due to the condition
(\ref{CONDITION}), such extra contributions become negligible in the
TDL.  Note that the analogous of Eq. (\ref{Ktc}) holds also for the
random variable $Q_{t}$, namely,
\begin{align}
  \label{Qtc}
  \mathbb{P}_t(\widetilde{\bm{n}}; Q_t\geq 1)
  \leq 1-\mathbb{P}_t(\widetilde{\bm{n}}; K_t=0).
\end{align}
Therefore, we can now use Eq.~(\ref{Kte}) and get
\begin{align}
  \label{Ktg}
  &\sum_{\widetilde{\bm{n}}\in
    \mathcal{A}^{(\mathrm{out})}(\widehat{\bm{n}})}
    \int_0^t ds \Gamma e^{-\Gamma A(\widehat{\bm{n}})s}
    \left(1- e^{-\sup_{\widetilde{\bm{n}}\in
    \widetilde{\partial}}\Gamma A^{(\mathrm{out})}
    (\widetilde{\bm{n}})(t-s)}\right)
    \nonumber\\
  &\qquad=
    \frac{A^{(\mathrm{out})}(\widehat{\bm{n}})}{A(\widehat{\bm{n}})}
    \left(1- e^{-\Gamma A(\widehat{\bm{n}})t}\right)
    \nonumber \\
  &\qquad\quad-
    A^{(\mathrm{out})}(\widehat{\bm{n}})
    e^{-\sup_{\widetilde{\bm{n}}\in \widetilde{\partial}}
    \Gamma A^{(\mathrm{out})}(\widetilde{\bm{n}})t}
    \frac{\left(1- e^{-\Gamma \left[A(\widehat{\bm{n}})-
    \sup_{\widetilde{\bm{n}}\in \widetilde{\partial}}
    A^{(\mathrm{out})}(\widetilde{\bm{n}})\right]t}\right)}
    {A(\widehat{\bm{n}})-\sup_{\widetilde{\bm{n}}\in \widetilde{\partial}}
    A^{(\mathrm{out})}(\widetilde{\bm{n}})}
    \nonumber \\
  &\qquad\leq 
    \frac{A^{(\mathrm{out})}(\widehat{\bm{n}})}{A(\widehat{\bm{n}})}
    -\frac{A^{(\mathrm{out})}(\widehat{\bm{n}})
    e^{-\sup_{\widetilde{\bm{n}}\in \widetilde{\partial}}
    \Gamma A^{(\mathrm{out})}(\widetilde{\bm{n}})t}}
    {A(\widehat{\bm{n}})-\sup_{\widetilde{\bm{n}}\in
    \widetilde{\partial}}A^{(\mathrm{out})}(\widetilde{\bm{n}})}
    \nonumber \\
  &\qquad\leq
    1- e^{-\sup_{\widetilde{\bm{n}}\in \widetilde{\partial}}
    \Gamma A^{(\mathrm{out})}(\widetilde{\bm{n}})t},     
\end{align}
where, for the last inequality, we have used
$A(\widehat{\bm{n}})\geq A^{(\mathrm{out})}(\widehat{\bm{n}})$, valid
for any configuration.
In conclusion, also for the configurations in $\widehat{\mathcal{H}}$
we have
\begin{align}
  \label{Kth}
  \sup_{\widehat{\bm{n}}} \mathbb{P}_t(\widehat{\bm{n}}; L_t\geq 1)
  \leq 1- e^{-\sup_{\widetilde{\bm{n}}\in \widetilde{\partial}}
  \Gamma A^{(\mathrm{out})}(\widetilde{\bm{n}})t}.
\end{align}
Equations~(\ref{Kte}) and (\ref{Kth}) show that what matters is always
the smallest border crossing-rate determined by
$\sup_{\widetilde{\bm{n}}\in \widetilde{\partial}}
A^{(\mathrm{out})}(\widetilde{\bm{n}})$. This concludes the proof of
Eq.~(\ref{MATRIX}).

\section{Equivalence of the conditions
  $A_\mathrm{norm}^{(\mathrm{out})}/N \to 0$ and
  $\dim{\mathcal{H}}_{\mathrm{cond}}/\dim{\mathcal{H}}\to 0$}

We have stated that Eqs.~(\ref{TH_QPTR}) are valid under the condition
$\sup_{\widetilde{\bm{n}}} A^{(\mathrm{out})}(\widetilde{\bm{n}})/N
\to 0$.  On the other hand, from~\cite{QPTA} we know that
Eqs.~(\ref{TH_QPTR}) at $T=0$ are valid under the condition
$\dim{\mathcal{H}}_{\mathrm{cond}}/\dim{\mathcal{H}}\to 0$.  It is
hence important to establish a relation between these two apparently
independent conditions.

We recall that the matrix elements of the hopping operator $K$ induce
in $\mathcal{H}$ a graph with $\dim{\mathcal{H}}$ nodes represented by
the configurations, where the degree of a configuration $\bm{n}$ is
given by its number of links $A(\bm{n})$.  In the following, we shall
focus only on regular qubit systems of $N$ qubits so that
$\dim{\mathcal{H}}=2^N$, and ``regular'' here means that the hopping
operator $K$ is made by the usual sum of $N$ single-flip operators, so
that $A(\bm{n}) \equiv N$.  Note that, since
$A(\bm{n})/\dim{\mathcal{H}}\to 0$, the graph associated to
$\mathcal{H}$ is a regular sparse graph~\cite{Bollobas}.

\begin{figure}[h]
  \begin{center}
    {\includegraphics[width=0.5\textwidth,clip]{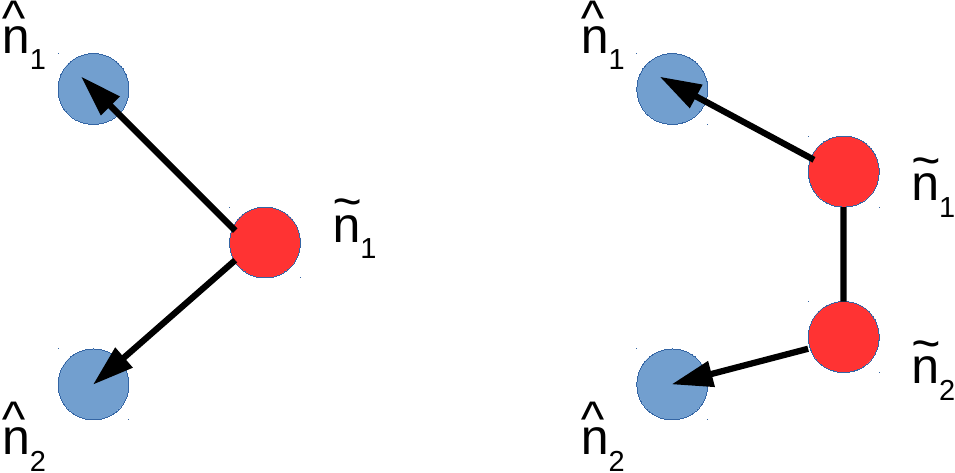}}
    \caption{Left Panel: an example of two configurations of
      $\widehat{\mathcal{H}}$ which differ by the status of two spins.
      In this case there exists a configuration of
      $\widetilde{\mathcal{H}}$ connected to the two configurations of
      $\widehat{\mathcal{H}}$.  Right Panel: an example of two
      configurations of $\widehat{\mathcal{H}}$ which differ by the
      status of three spins.  In this case there is no configuration
      of $\widetilde{\mathcal{H}}$ connected directly to both the two
      configurations of $\widehat{\mathcal{H}}$.}
    \label{fig3}
  \end{center}
\end{figure}

Let us first consider the Grover model. This model is characterized by
the fact that there exist only two possible values of the potential,
$V=-JN$ e $V=0$, and that the former is realized by just one
configuration (for example the one in which all the spins are up) so
that $\dim{\mathcal{H}}_{\mathrm{cond}}=1$. For this model we have
$A^{(\mathrm{out})}(\widetilde{\bm{n}}) \leq 1$ and also
$\dim{\mathcal{H}}_{\mathrm{cond}}/\dim{\mathcal{H}}=1/2^N\to 0$.  We
can generalize the Grover model by allowing the value $V=-JN$ to be
$\dim{\mathcal{H}}_{\mathrm{cond}}>1$ degenerate provided that we
still have $\dim{\mathcal{H}}_{\mathrm{cond}}/\dim{\mathcal{H}}\to
0$. It is clear that, as far as the
$\dim{\mathcal{H}}_{\mathrm{cond}}$ configurations associated to
$V=-JN$ are sufficiently separated, we keep having
$A^{(\mathrm{out})}(\widetilde{\bm{n}}) \leq 1$. More precisely, it is
easy to see that, as far as the $\dim{\mathcal{H}}_{\mathrm{cond}}$
configurations associated to $V=-JN$ differ for the values of at least
three spins (i.e., in the graph, the configurations of
$\widehat{\mathcal{H}}$ are at least three links far apart among each
other), we still have $A^{(\mathrm{out})}(\widetilde{\bm{n}}) \leq 1$
for any $\widetilde{\bm{n}}$. This condition is illustrated in
Fig.~\ref{fig3}.  However, it should be clear that the condition
$\dim{\mathcal{H}}_{\mathrm{cond}}/\dim{\mathcal{H}}\to 0$ alone in
general does not imply the condition
$A^{(\mathrm{out})}(\widetilde{\bm{n}})=O(1)$. As a counter-example,
if we define $\widehat{\mathcal{H}}$ as the set of the $N$
configurations first neighbors of a given one, $\widetilde{\bm{n}}$,
we see that by construction
$\dim{\mathcal{H}}_{\mathrm{cond}}/\dim{\mathcal{H}}\to 0$ but now
$A^{(\mathrm{out})}(\widetilde{\bm{n}})=N$ (indeed, here the
$\dim{\mathcal{H}}_{\mathrm{cond}}$ configurations associated to
$V=-JN$ differ for the direction of two spins).

\begin{figure}[h]
  \begin{center}
    {\includegraphics[width=0.5\textwidth,clip]{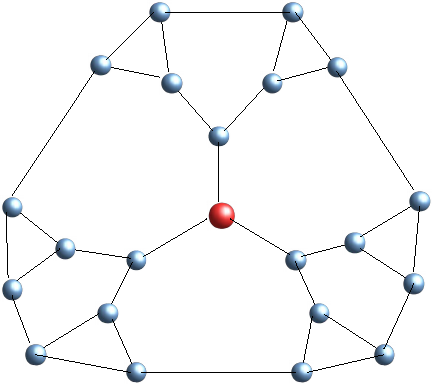}}
    \caption{A regular graph of degree $A=3$ drawn from the
      perspective of the ``central'' red node. The subgraphs having a
      boundary at the distances $l=1$ or $l=2$ from the central node,
      i.e., those obtained by removing all the nodes at distance
      larger than $l$ as well as all the links emanating from these
      removed nodes, are Cayley trees of degree $A=3$ (except for the
      boundary, where the nodes have degree 1). However, the
      complements of these subgraphs are not trees. In particular, the
      complement of the case $l=2$ is a regular graph of degree 2.}
    \label{fig2}
  \end{center}
\end{figure}

The above counter example, however, is rather nonphysical as it does
not take into account how the structure of a physical potential
operator $V$ acts on the definition of
$\mathcal{H}_{\mathrm{cond}}$. The definition of
$\mathcal{H}_{\mathrm{cond}}$ is in principle arbitrary but the most
interesting cases are those in which $\mathcal{H}_{\mathrm{cond}}$ is
defined directly from the structure of the operator $V$. The idea is
to define $\mathcal{H}_{\mathrm{cond}}$ through the configurations
$\bm{n}$ having potential levels
$V(\bm{n})=\braket{\bm{n}}{V}{\bm{n}}$ not larger than some threshold
value $\max V_{\mathrm{cond}}$, namely,
$\widehat{\mathcal{H}} = \{\bm{n}:~V(\bm{n})\leq \max
V_{\mathrm{cond}}\}$.  For given $N$, if $V$ has some physical origin,
$\dim{\mathcal{H}}_{\mathrm{cond}}$ is expected to be a fast growing
function of $\max V_{\mathrm{cond}}$, typically exponential.  Notice,
however, that this assumption holds true for not too large values of
$\max V_{\mathrm{cond}}$, being $\dim{\mathcal{H}}_{\mathrm{cond}}$
limited by $\dim{\mathcal{H}}$.  In fact, it holds true as far as
$\dim{\mathcal{H}}_{\mathrm{cond}}/\dim{\mathcal{H}}\ll 1$.  As a
consequence, if
$\dim{\mathcal{H}}_{\mathrm{cond}}/\dim{\mathcal{H}}\ll 1$, the
subgraph induced by $K$ on the set $\widehat{\mathcal{H}}$, can
effectively be treated as a regular Cayley tree of size
$\dim{\mathcal{H}}_{\mathrm{cond}}$ and degree $N$, i.e., a finite
graph without loops where each node has degree $N$, except for its
boundary, where the nodes have degree 1. This assumption corresponds
to the usual tree-like approximation that holds true locally in many
sparse graphs.  By contrast, the subgraph induced by $K$ on the set
$\widetilde{\mathcal{H}}$ cannot be treated as a tree. If fact, we
have to take into account that the total graph induced by $K$ in
$\mathcal{H}$, is a regular graph without an actual boundary; it is
not a tree. As a consequence, we see that the complement of any tree
in the total graph, and therefore also in the subgraph induced by $K$
on the set $\widetilde{\mathcal{H}}$, cannot be treated as a tree
either, see Fig.~\ref{fig2} for an illustrative example.  More
precisely, in the graphs associated to $\mathcal{H}$ and
$\widetilde{\mathcal{H}}$ there are loops whose shortest length $l$ is
of the order $l=\log(\dim{\mathcal{H}})/\log(N)$.

\begin{figure}[h]
  \begin{center}
    {\includegraphics[width=0.45\textwidth,clip]{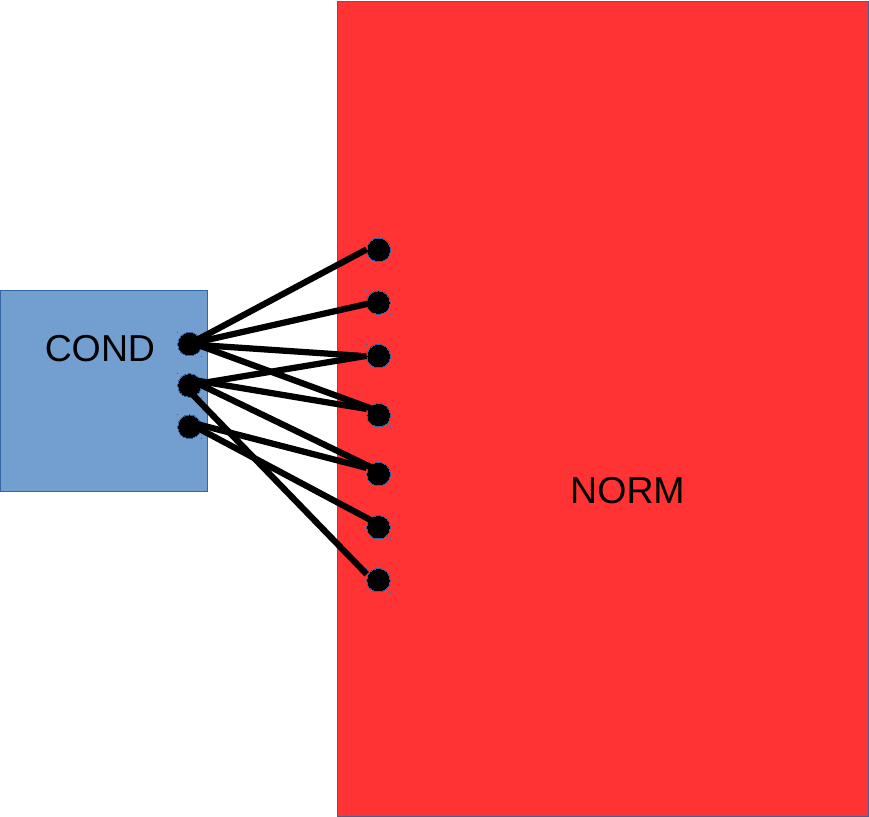}}
    \caption{A schematic example of a cond-norm partition where the
      nodes (configurations) on the two boundaries,
      $\widehat{\partial}$ and $\widetilde{\partial}$, are put in
      evidence near the two adjacent sides. Also the links connecting
      the two boundaries are put in evidence. The
      $|\widehat{\partial}|=3$ nodes in $\widehat{\partial}$ have
      degrees 4, 4, 2, while the $|\widetilde{\partial}|=7$ nodes in
      $\widetilde{\partial}$ have degrees 1, 1, 2, 2, 2, 1, 1.  We can
      read the total number of links connecting the two boundaries
      from cond to norm as
      $4+4+2=10=\overline{A}_{\mathrm{cond}}^{(\mathrm{out})}
      |\widehat{\partial}|$, or else from norm to cond as
      $1+1+2+2+2+1+1=10=\overline{A}_{\mathrm{norm}}^{(\mathrm{out})}
      |\widetilde{\partial}|$, where
      $\overline{A}_{\mathrm{cond}}^{(\mathrm{out})}$ and
      $\overline{A}_{\mathrm{norm}}^{(\mathrm{out})}$ are the mean
      numbers of the outgoing links of the two partitions. In this
      example we have
      $\overline{A}_{\mathrm{cond}}^{(\mathrm{out})}=10/3$ and
      $\overline{A}_{\mathrm{norm}}^{(\mathrm{out})}=10/7$.  }
    \label{fig4b}
  \end{center}
\end{figure}

As is known, one peculiar feature of the Cayley tree is the fact that
its boundary constitutes a finite portion of its total number nodes,
see for example~\cite{CTBL}.  Moreover, we have to take into account
the constraint that the total number of outgoing links from
$\widetilde{\mathcal{H}}$ to $\widehat{\mathcal{H}}$ must be equal to
the total number of outgoing links from $\widehat{\mathcal{H}}$ to
$\widetilde{\mathcal{H}}$.  By making use of the mean numbers of
outgoing links $\overline{A}_{\mathrm{cond}}^{(\mathrm{out})}$ and
$\overline{A}_{\mathrm{norm}}^{(\mathrm{out})}$, along the boundaries
$\widehat{\partial}$ and $\widetilde{\partial}$, respectively, we have
(see Fig.~\ref{fig4b} for an illustrative example)
\begin{align}
  \label{AAA}
  \overline{A}_{\mathrm{cond}}^{(\mathrm{out})}
  ~|\widehat{\partial}|
  =
  \overline{A}_{\mathrm{norm}}^{(\mathrm{out})}
  ~|\widetilde{\partial}|,
\end{align}
which, if we call $\alpha_{\mathrm{cond}}$ the coefficient providing
$|\widehat{\partial}|=\alpha_{\mathrm{cond}}
\dim{\mathcal{H}}_{\mathrm{cond}}$ and use
$|\widetilde{\partial}|\leq
\dim{\mathcal{H}}-\dim{\mathcal{H}}_{\mathrm{cond}}$, gives
\begin{align}
  \label{AAA1}
  \overline{A}_{\mathrm{cond}}^{(\mathrm{out})}
  \alpha_{\mathrm{cond}}\dim{\mathcal{H}}_{\mathrm{cond}}\leq 
  \overline{A}_{\mathrm{norm}}^{(\mathrm{out})}
  \left(\dim{\mathcal{H}}-\dim{\mathcal{H}}_{\mathrm{cond}}\right)
  \leq \overline{A}_{\mathrm{norm}}^{(\mathrm{out})}\dim{\mathcal{H}}.
\end{align}
For a regular Cayley tree of degree $N$ it is easy to see that
$\alpha_{\mathrm{cond}} \to 1^{-}$ so that Eq.~(\ref{AAA1}) gives
\begin{align}
  \label{AAA2}
  \frac{\dim{\mathcal{H}}_{\mathrm{cond}}}{\dim{\mathcal{H}}} \leq
  \frac{\overline{A}_{\mathrm{norm}}^{(\mathrm{out})}}
  {\overline{A}_{\mathrm{cond}}^{(\mathrm{out})}}
\end{align}
Finally, since $\overline{A}_{\mathrm{cond}}^{(\mathrm{out})}=O(N)$,
Eq.~(\ref{AAA2}) proves that the condition
$\overline{A}_{\mathrm{norm}}^{(\mathrm{out})}/N\to 0$ implies the
condition $\dim{\mathcal{H}}_{\mathrm{cond}}/\dim{\mathcal{H}}\to 0 $.

The above Eq.~(\ref{AAA2}) is exact but it does not allow to prove the
converse. Nevertheless, if we take into account the exponential growth
with $N$ of $\dim{\mathcal{H}}$, holding for most of the systems of
interest, Eq.~(\ref{AAA2}) leads us to make the following ansatz
\begin{align}
  \label{AAA3}
  \frac{\overline{A}_{\mathrm{norm}}^{(\mathrm{out})}}
  {\overline{A}_{\mathrm{cond}}^{(\mathrm{out})}}
  \sim -1/ \log
  \left(\frac{\dim{\mathcal{H}}_{\mathrm{cond}}}{\dim{\mathcal{H}}}\right),
\end{align}
which in turn implies that
$\dim{\mathcal{H}}_{\mathrm{cond}}/\dim{\mathcal{H}}\to 0 $ if and
only if $\overline{A}_{\mathrm{norm}}^{(\mathrm{out})}/N\to 0$.

The ansatz (\ref{AAA3}) is compatible with Eq.~(\ref{AAA2}) and is
clearly satisfied in the case of the Grover model and its
generalizations.  To make concrete the ansatz with a more physical
example, let us consider the interaction potential of the
one-dimensional Ising model with periodic boundary conditions. If we
represent the configurations by products of single spin states along
the $z$ axis,
$|\bm{n}\rangle = |\sigma^z_1\rangle \otimes \cdots \otimes
|\sigma^z_{N}\rangle $, with $\sigma^z_i=\pm 1$, $i=1,\dots,N$, we
have
\begin{align}
  \label{Ising}
  \langle \bm{n}|V|\bm{n}\rangle = V(\bm{n})=
  -J \sum_{i=1}^N\sigma^z_i\sigma^z_{i+1}.
\end{align}
We assume $J>0$.  We are free to define $\widehat{\mathcal{H}}$ (and
then $\mathcal{H}_{\mathrm{cond}}=\Span \{\widehat{\mathcal{H}}\}$) in
several ways, and we want to see to what extent the conditions
$\dim{\mathcal{H}}_{\mathrm{cond}}/\dim{\mathcal{H}}\to 0 $ and
$\overline{A}_{\mathrm{norm}}^{(\mathrm{out})}/N\to 0$ are equivalent.
We can start by including in $\widehat{\mathcal{H}}$ the two lowest
ground states with all parallel spins. Then we can enlarge
$\widehat{\mathcal{H}}$ by including all the states in which one spin
is reversed with respect to all the other $N-1$ parallel ones and so
on.  Alternatively and more effectively, we can characterize any
configuration by the number of cuts $q$ in it, where a cut is present
if, reading the sequence of the pointers $\sigma^z_i$ for example from
left to right, we meet an inversion. In terms of $q$ Eq.~(\ref{Ising})
reads (we can have at most $N-1$ number of cuts and we start by
considering the two ground states in which all the spins are parallel)
\begin{align}
  \label{Ising1}
  V_q=-JN+2Jq, \qquad D(q)=2\binom{N-1}{q}, \qquad q=0,\dots,N-1,
\end{align}
where $D(q)$ is the number of configurations $\bm{n}$ having potential
$V(\bm{n})=V_q$ .  We define $\widehat{\mathcal{H}}$ by introducing a
threshold $\max V_{\mathrm{cond}}$ as the maximum allowed potential
value of its configurations. If we choose
$\max V_{\mathrm{cond}}=V_k$, we have
$\widehat{\mathcal{H}}=\widehat{\mathcal{H}}(k)$ with
\begin{align}
  \label{Ising2}
  \widehat{\mathcal{H}}(k) =
  \left\{\bm{n}:~V(\bm{n})\leq V_k \right\},
  \qquad \dim{\mathcal{H}}_{\mathrm{cond}}=2\sum_{q=0}^k \binom{N-1}{q}.
\end{align}
By recalling that $\binom{N}{k}/2^N$ tends, for $N\to\infty$, to a
Dirac delta function centered at $k=N/2$, we see that
\begin{align}
  \label{Ising3}
  \frac{\dim{\mathcal{H}}_{\mathrm{cond}}}{\dim{\mathcal{H}}} \to 0
  \qquad \mathrm{as~soon~as}
  \qquad \frac{k}{N}< \frac{1}{2}. 
\end{align}
On the other hand, we can verify that the condition on
$A_{\mathrm{norm}}^{(\mathrm{out})}$ is satisfied whenever $k/N <1/2$
as follows. Given $k$, let us consider the boundary of
$\widetilde{\mathcal{H}}$
\begin{align}
  \label{Ising4}
  \widetilde{\partial}=\left\{\bm{n}:~  V(\bm{n})=V_{k+1}\right\}.
\end{align}
Given $\widetilde{\bm{n}}\in\widetilde{\partial}$, by inverting one of
its spins located at a cut, the cut will be either shifted or removed,
leaving the potential unchanged or lowered by $2J$ (and then entering
$\widehat{\mathcal{H}}$), respectively.  It is instructive to consider
the two opposite regimes: $k$ very small, and $k$ very large. The
former regime occurs when $k\ll N$ as when a few isolated spins are
antiparallel to the others. In this case we have
$A_{\mathrm{norm}}^{(\mathrm{out})}(\widetilde{\bm{n}})=O(k)$.  The
other regime occurs when there are nearly half spins up and half spins
down, i.e., when $k\sim N/2$, where
$\dim{\mathcal{H}}_{\mathrm{cond}}/\dim{\mathcal{H}}=O(1)$, and here
we have $A_{\mathrm{norm}}^{(\mathrm{out})}(\widetilde{\bm{n}})=O(N)$.
In the intermediate regime we have
$A_{\mathrm{norm}}^{(\mathrm{out})}(\widetilde{\bm{n}})=o(N)$, i.e.,
non-extensive.  This example shows that the conditions
$A_\mathrm{norm}^{(\mathrm{out})}/N \to 0$ and (\ref{Ising3}) are
essentially equivalent and that the ansatz (\ref{AAA3}) is realized
with
$\dim{\mathcal{H}}_{\mathrm{cond}}/\dim{\mathcal{H}} \sim \exp(k-N)$.
However, we warn that, as we have shown in~\cite{QPTA}, in the case of
the Ising model, Eq.~(\ref{QPT0}) has no solution, whatever $k$, so
that our theory turns out to be not useful in such a case, as it
always occurs when the QPT is second-order. Yet, the above picture is
very general and can be similarly applied to many models, as in the
particularly important case of interacting fermions~\cite{WC_QPT}
(where the QPT is first-order). We have directly checked that in all
these models the condition
$A_{\mathrm{norm}}^{(\mathrm{out})}(\widetilde{\bm{n}})/N \to 0$ turns
out to be satisfied and that the ansatz (\ref{AAA3}) holds true.

\ack
  Grant CNPq 307622/2018-5 - PQ (Brazil) is acknowledged.
  M.~O. thanks the Istituto Nazionale di Fisica Nucleare, Sezione di
  Roma 1, and the Department of Physics of Sapienza University of Rome
  for financial support and hospitality.

\section*{References}


\begin{thebibliography}{10}
  \expandafter\ifx\csname url\endcsname\relax \def\url#1{{\tt #1}}\fi
  \expandafter\ifx\csname urlprefix\endcsname\relax\def\urlprefix{URL
  }\fi \providecommand{\eprint}[2][]{\url{#2}}


\bibitem{SGCS} S.~L.~Sondhi, S.~M.~Girvin, J.~P.~Carini, and
  D.~Shahar, ``Continuous quantum phase transitions'',
  Rev. Mod. Phys. \textbf{69}, 315 (1997).
  
\bibitem{KB} T.~R.~Kirkpatrick and D.~Belitz, ``Quantum phase
  transitions in electronic systems'', in \textit{Electron
    Correlations in the Solid State} ed. by N.~H.~March, (Imperial
  College Press, London 1999).

\bibitem{Vojta} T.~Vojta, ``Quantum phase transitions in electronic
  systems'', Ann. Phys. (Leipzig) \textbf{9}, 403 (2000).

\bibitem{Sachdev} S.~Sachdev, \textit{Quantum Phase Transitions}
  (Cambridge University Press, Cambridge 2000).

\bibitem{Carr} L.~D.~Carr, ``Understanding Quantum Phase
  Transitions'', (CRC Press, Taylor \& Francis 2010).

\bibitem{Plastino} A.~Plastino and E.~M.~F.~Curado, ``Finite
  temperature approach to quantum phase transitions'', International
  Journal of Bifurcation and Chaos \textbf{20}, 397 (2010).

\bibitem{PRL} M.~Ostilli and C.~Presilla, ``Finite temperature quantum
  condensations in the space of states'', arXiv:2203.05803 (2022).
  
\bibitem{Sebenik} A.~B.~Finilla, M.~A.~Gomez, C.~Sebenik, and
  D.~J.~Doll, ``Quantum annealing: A new method for minimizing
  multidimensional functions'', Chem. Phys. Lett. \textbf{219}, 343
  (1994).

\bibitem{Nishimori} T.~Kadowaki and H.~Nishimori, ``Quantum annealing
  in the transverse Ising model'', Phys. Rev. E \textbf{58}, 5355
  (1998).
    
\bibitem{Santoro} G.~E.~Santoro and E.~Tosatti, ``Optimization using
  quantum mechanics: quantum annealing through adiabatic evolution'',
  J. Phys. A \textbf{39}, R393 (2006).

\bibitem{QPTA} M.~Ostilli and C.~Presilla, ``First-order quantum phase
  transitions as condensations in the space of states'', J. Phys. A:
  Math. Theor. \textbf{54}, 055005 (2021).
  
\bibitem{Grover} L.~K.~Grover, ``A fast quantum-mechanical search
  algorithm for database search'', Proceedings, 28th Annual ACM
  Symposium on the Theory of Computing (STOC), May 1996, pages
  212-219; ``Quantum Mechanics Helps in Searching for a Needle in a
  Haystack'', Phys. Rev. Lett. \textbf{79}, 325 (1996); ``From
  Schr\"odinger's equation to the quantum search algorithm'',
  Am. J. Phys. \textbf{69}, 769 (2001).
  
\bibitem{Farhi.Gutmann} E.~Farhi and S.~Gutmann, ``Analog analogue of
  a digital quantum computation'', Phys. Rev. A \textbf{57}, 2403
  (1998).

\bibitem{Roland.Cerf} J.~Roland and N.~J.~Cerf, ``Quantum search by
  local adiabatic evolution'', Phys. Rev. A, \textbf{65}, 042308
  (2002).
 
\bibitem{Jorg:2008} T.~J\"org, F.~Krzakala, J.~Kurchan, and
  A.~C.~Maggs, ``Simple glass models and their quantum annealing'',
  Phys. Rev. Lett. \textbf{101}, 147204 (2008).

\bibitem{Jorg:2010} T.~J\"org, F.~Krzakala, J.~Kurchan, A.~C.~Maggs,
  and J.~Pujos, ``Energy gaps in quantum first-order mean-field–like
  transitions: The problems that quantum annealing cannot solve'',
  Europhysics Letters \textbf{89}, 40004 (2010).

\bibitem{WC_QPT} M.~Ostilli and C.~Presilla, ``Wigner crystallization
  of electrons in a one-dimensional lattice: a condensation in the
  space of states'', Phys. Rev. Lett. \textbf{127}, 040601 (2021).

\bibitem{BPDAJL} M.~Beccaria, C.~Presilla, G.~F.~De Angelis, and
  G.~Jona-Lasinio, ``An exact representation of the fermion dynamics
  in terms of Poisson processes and its connection with Monte Carlo
  algorithms'', Europhys. Lett. {\bf 48}, 243 (1999).

\bibitem{AABEB} M.~Aparicio Alcalde, M.~Bucher, C.~Emary, and
  T.~Brandes, ``Thermal phase transitions for Dicke-type models in the
  ultrastrong-coupling limit'', Phys. Rev. E \textbf{86}, 012101
  (2012).

\bibitem{Continentino} M.~A.~Continentino and A.~S.~Ferreira,
  ``First-order quantum phase transitions'', J. Magnetism and Magnetic
  Materials \textbf{310}, 828 (2007).

\bibitem{Pelissetto} M.~Campostrini, J.~Nespolo, A.~Pelissetto, and
  E.~Vicari, ``Finite-Size Scaling at First-Order Quantum
  Transitions'', Phys. Rev. Lett. \textbf{113}, 070402 (2014).
  
\bibitem{Bollobas} B.~Bollob\'as, ``Random Graphs'' (2nd ed.),
  (Cambridge University Press 2001).
  
\bibitem{CTBL} M.~Ostilli, ``Cayley Trees and Bethe Lattices: A
  concise analysis for mathematicians and physicists'', Physica A,
  \textbf{391}, 3417 (2012).
  
\end{thebibliography}

\providecommand{\newblock}{}


\end{document}